\documentclass[12pt]{iopart}

\usepackage{iopams}  
\newcommand{\be}{\begin{equation}}
\newcommand{\ee}{\end{equation}}
\newcommand{\bea}{\begin{eqnarray}}
\newcommand{\eea}{\end{eqnarray}}
\usepackage{float}
\usepackage{graphicx}
\usepackage{xcolor}
\usepackage{bigints}
\usepackage{amsmath}
\usepackage{calrsfs}

\begin{document}

\title{Kink-like Solitons in Quantum Droplet}

\author{Aradhya Shukla }

\address{Department of Physical Sciences, Indian Institute of Science Education and Research
Kolkata, Mohanpur-741246, West Bengal, India}
\ead{aradhya02@iiserkol.ac.in}

\author{ Neeraj }

\address{Department of Physical Sciences, Indian Institute of Science Education and Research
Kolkata, Mohanpur-741246, West Bengal, India}
\ead{n15rs071@iiserkol.ac.in}

\author{ Prasanta K. Panigrahi\footnote{Corresponding author} }
\address{Department of Physical Sciences, Indian Institute of Science Education and Research
Kolkata, Mohanpur-741246, West Bengal, India}
\ead{pprasanta@iiserkol.ac.in}
\vspace{10pt}

\begin{abstract}
Solitonic excitations of the one-dimensional quantum droplets are obtained, which smoothly connect vacuum with the flat-top droplet, akin to compactons in classical liquids. These solitons are of the kink type, necessarily residing on a constant pedestal, determined by the mean-field repulsion and beyond mean field quantum correction
 and having exactly one-third of the uniform condensate amplitude. Akin to the kinks, the propagating modes occur in pairs and are phase-locked with the background. The lowest chemical potential and condensate amplitude at the flat-top boundary matches with the self-trapped quantum droplet. More general excitations of analogous kind are obtained through the M{\"o}bius transform, which connect the required solutions to elliptic functions in general.
 \end{abstract}

%
%
%
%
%

\section{Introduction}

Solitons are ubiquitous in one-dimensional non-linear dynamical systems \cite{ drezin} and have manifested in diverse physical systems \cite{sol, bis, poly_acy}, ranging from polyacetylene \cite{ pkp_poly} to optical fibres \cite{has, GPagr} and Bose-Einstein condensates (BECs) in cigar shaped BEC \cite{pethick, kev}. 
Characteristically, they take the form of dark \cite{densch, busch}, grey \cite{sh} and bright \cite{kh, mcd, aspect} solitons, respectively in the repulsive and attractive interaction regimes of BEC. Their existence and stability depend on the balancing effect of dispersion with non-linearity. 
On the other hand, the kink and anti-kink are solitonic solutions of the $\lambda \phi^4$-theory in one dimension  \cite{rajaram}, owing their stability to their topological charges \cite{um, vivek}. They interpolate between the two degenerate vacuua in the broken parity phase \cite{man}, while passing through the normal phase, where the order parameter vanishes. The analogous solutions for the complex order parameter are the previously mentioned, dark  and grey  solitons  of the non-linear Schr\"{o}dinger equation (NLSE), describing the mean-field dynamics of the one dimensional BEC. Akin to the kink and anti-kink, these extended objects asymptotically connect points on the degenerate vacuum manifold of the broken global $U(1)$ symmetry phase of BEC \cite{pethick, stringary}. The grey soliton is the well-known Lieb mode, first obtained in the second-quantized model \cite{lieb}, and subsequently found  as an exact solution of the NLSE \cite{Kul, 1r27, jack}, which connects points in the vacuum  manifold without passing through the normal phase. It has been observed in BEC \cite{sh, becker, romero}, and later as  excitations in water body \cite{sol_ocn, Ca} and other physical systems \cite{DNC, GBC}.  Generically, these excitations are composed of the hyperbolic tangent function, which asymptotically takes positive and negative values. The well-known bright soliton has a characteristic profile in the form of the hyperbolic secant function, vanishing at spatial infinity \cite{kiv}. 



Recently, a new form of quantum matter, the  quantum droplet \cite{1r5, 1r5_ptr}, has been identified in BEC, after taking into consideration beyond mean-field (BMF) Lee-Huang-Yang (LHY) quantum correction \cite{lee}.The droplets are self-bound and exist in free space, recently observed experimentally in a number of systems like Bose-Bose (BB) mixture \cite{1r14, 1r15, 1r16} and dipolar BEC \cite{3r1}. 
 Experimentally, variation of the coupling from  weak to strong attractive domains yields a transition from expanding to localized state, which agrees with theoretical BMF predictions.  The equilibrium properties of quantum droplets, e.g., size, critical number density and binding energy, have been found in agreement with the theoretical predictions \cite{FSM1, FSM2}.
 The dimension crossovers from 1D$\rightarrow$3D  \cite{Lavoine} and 2D$\rightarrow$3D in binary BECs have also been  analysed  \cite{malo}. Droplet  with constituents having electric and magnetic moments have been explored \cite{cm}, where it is observed that the droplets experience a crossover from the cigar to pancake shapes in terms of relative dipole orientations. 
 A new type of droplet in binary magnetic gases has been recently found, where single component self-bound droplet can couple with other magnetic components, not in the droplet phase \cite{sm}.
The study of collective excitations in the form of Goldstone modes, corresponding to the spontaneously broken internal and  translational symmetries, is the subject of many recent investigations \cite{malo2, pfou, petrov, pfou1}.


It is well-established that the BMF effect emerges from the zero-point energy summation of the Bogoliubov modes and depends on the dimensions of the system \cite{1r5, 1r5_ptr, 1r16}. The BMF correction, in three-dimensions is attractive and scales as $n^{5/2}$ ($n$ being the number density), whereas in one-dimension,  scaling is proportional to $n^{3/2}$ and repulsive in nature \cite{1r16, malo}. The corresponding mean field equation is the amended NLSE, having cubic and quadratic non-linearities.
In one-dimension, exact self-bound droplet solution has been obtained by  Petrov and Astrakharchik \cite{1r5_ptr}, revealing its characteristic  flat-top nature.   As is well-known, modulation instability (MI) dictates the stability of the propagating modes in non-linear systems. Recently, the growth rate of MI for one-dimensional quantum droplet and BB mixture have been investigated \cite{pkp, kare}.  The  possible MI of this system has been studied, with and without spin-orbit coupling, indicating the parameter domain where the instability of propagating plane waves may lead to solitonic excitation. The parameter domain beyond MI is conducive for generation of soliton and soliton trains \cite{hulet1}.

Here, we explicitly demonstrate kink-like solitons in the droplet regime, similar in structure to dark and grey solitons, but also showing  significant differences. These solitons necessarily require the presence of a constant background, which  is exactly one-third of the uniform condensate amplitude: $(\frac{\sqrt{2M}}{\pi \hbar})g^{3/2}/\delta g$. They smoothly connect the normal zero-condensate vacuum with the droplet configuration and occur in pairs, similar to kink/anti-kinks in the $\lambda\phi^4$-theory \cite{rajaram}. However, these excitations asymptotically approach vanishing condensate density only at one end, unlike kink/anti-kinks and dark/grey solitons, which connect the degenerate vacua at both the asymptotic domain.  They are similar to compactons \cite{5r2}, manifesting in the liquid-normal material boundary, having strictly compact support with vanishing derivatives. However, the quantum liquid solitons smoothly interpolate between the droplet and the normal phase. The fact that the droplet has a broad flat density profile makes it plausible that these solitonic excitations may appear at the boundary of the droplets, wherein the condensate phase smoothly joins with the normal phase.

The paper is organized as follows. In Sec. II, theory of quantum droplets is briefly described leading to the amended Gross-Pitaevskii (GP) equation in the form of NLSE with quadratic coupling, governing its dynamics in one-dimension. Sec. III is devoted to obtain the consistency relations for the kink-like solution with background  and obtain its stability criteria.  It is to be mentioned that the consistency conditions lead to a bounded below chemical potential, which is identical to the lowest chemical potential of droplets. In Sec. IV, we consider a more general exact solution of a kink-like behavior and find the inter-connection between the background, amplitude and healing length of the  solitonic profile. In Sec. V, using the suitable background subtraction, we obtain the ground state energy and momentum for kink-like soliton. Finally, we conclude with the summary of results and future directions for investigation.  



\section{Theory of Quantum Droplets in One Dimension}   

Quantum droplets have been observed by exploiting  the Bose-Bose mixture of ultracold atoms. The formation of quantum droplets are the result of balance between the MF and BMF interactions such that $0<\delta g \ll g$, where $g=g_{\uparrow \uparrow}=g_{\downarrow \downarrow}$, is the intra-particle interaction with each component having equal number of atoms, $n = n_{\uparrow} = n_{\downarrow}$ and $\delta g = g_{\uparrow \downarrow}+g$, being the inter-particle interaction.

It has been illustrated earlier that, in three-dimensions, MF and BMF interactions scale for  density is different : $E_{MF} \propto n^2$ and $E_{BMF} \propto n^{5/2}$, whereas, in one-dimension, energy of the effective one component BEC is,
\be \label{eqn:1}
   E_{1D} = \frac{\delta g n^2}{2}-\frac{2\sqrt{2m}\left(gn\right)^{3/2}}{3\pi \hbar},
 \ee
with $n_0 = 8g^3/(9 \pi^2 \delta g^2)$ being the equilibrium density and chemical potential $\mu_0 = -\delta g n_0/2$. 
 Thus, at certain density, MF and BMF effects compensate each other leading to the emergence of stable droplets. The negative chemical potential is essential  as it prevents self-evaporation of droplets. 

In the case of one-dimension, the mean field dynamical equation takes the form of amended GP equation:

\be \label{eqn:2}
    i \hbar \frac{\partial \Psi}{\partial t} =  -\frac{\hbar^2}{2m} \frac{\partial^2 \Psi}{\partial x^2}  +\delta g~ \left|\Psi\right|^2\Psi- \frac{\sqrt{2m}}{\pi \hbar}g^{3/2} \left|\Psi\right|\Psi,
\ee
characterized by cubic and quadratic non-linearities. This mean-field expression arises from an effective two-components BEC, with attractive inter-particle and intra-particle interaction in the region, $0<\delta g \ll g$.

We consider the transformation $\Psi(x, t) = \Phi(x-vt)\; exp\, \big[i(k\,x - \frac{\mu}{\hbar}\, t) \big]$  and substitute into the amended NLSE equation. Comparing the imaginary and real  parts lead to two coupled equations. First one is the continuity equation as given below:
\begin{equation}
\frac{\partial \Phi}{\partial t} + \frac {\hbar k}{m}\, \frac{\partial \Phi}{\partial X} = 0,
\label{eqn:3}
\end{equation}
with $X = x-vt$ and $v = \hbar k/ m.$ The second equation originating from the real part reads as,
\be
-\frac{\hbar^2}{2m} \frac{\partial^2 \Phi}{\partial X^2} + \delta g {|\Phi|}^2 \Phi -\frac{\sqrt{2m}}{\pi \hbar}g^{3/2} |\Phi| \Phi - \bar\mu \Phi = 0.\label{eqn:4}
\ee
 The term $\hbar^2 k^2/2m$, coming from the kinetic part of amended GP equation, has been absorbed into $\bar\mu = \mu  - \hbar^2 k^2/2m$. 
 Considering a constant solution  $\Phi(X)  = \sqrt{\sigma_0}$, Eq. (\ref{eqn:4}) leads to the following 
\be
\delta g\,\sigma_0- \frac{\sqrt{2m}}{\pi \hbar}g^{3/2}\, \sqrt{\sigma_0} - \mu = 0, \label{eqn:5}
\ee
which gives two distinct allowed values, 
\be
\sqrt{\sigma_0}_{\pm} = \alpha \Bigl[ 1 \pm \Big(1+ \frac{\mu}{\delta g\, \alpha^2} \Big)^{1/2}  \Bigr], \label{eqn:6}
\ee
with $\alpha = \big(\frac{\sqrt{2m}}{\pi \hbar}\big)\, g^{3/2}/2\delta g$. 
 It is noteworthy that, for a particular $\mu = -(\frac{4m}{9\pi^2 \hbar^2})\, g^3/\delta g,$ the resulting $\sqrt{\sigma_0}_+$ and $\sqrt{\sigma_0}_-$ are same whereas for any other $\mu$, the constant solution leads to two different  backgrounds,  which establishes its non-analytic nature. 
In the next section, we take the soliton solution having the kink-like behavior and illustrate its properties.
 

\section{Kink-like Soliton}   

It is to be noted that the amended NLSE has non-linearity, similar to the weak and strong coupling domains of BEC \cite{sal1}. Although the kink-type solutions exist in NLSE, the same has not been existed for the strong coupling. Therefore we consider a general ansatz having a propagating kink-like excitation, phase-locked with a non-zero background:
\be
     \Phi(X)=A+B~ \tanh\big( X/\xi \big),
    \label{ans1}
\ee
where $A, B$ are constants with $\xi$ being the healing length. Substituting the above in Eq. (\ref{eqn:4}), one gets the following relationships: 
\bea
&\frac{1}{\xi^2} = \Big(\frac{\delta g m}{\hbar^2}\Big) B^2, \qquad \quad A = \Big(\frac{\sqrt{2m}}{3\pi \hbar}\Big)\, \frac{g^{3/2}}{\delta g},\nonumber\\
&B = \pm\Big(\frac{\sqrt{2m}}{3\pi \hbar}\Big)\, \frac{g^{3/2}}{\delta g}, \qquad \frac{\hbar^2\,k^2}{2m} =  \mu +\Big(\frac{4\,m}{9\pi^2 \hbar^2}\Big)\, \frac{g^3}{\delta g},
  \label{eqn:7}
\eea
which illustrate that healing length is inversely proportional to soliton amplitude and controlled by the MF coupling. From the above equation, it is evident  that the solitons necessarily reside on a constant condensate background, which is exactly one-third of the uniform condensate amplitude. They occur in pairs, $B=\pm A$, vanishing asymptotically at one end, and connecting the normal vacuum with the quantum droplets located at the origin, with an appropriate translation of soliton profile. 
 This result is in agreement with the Petrov's flat bulk region for droplet \cite{1r5_ptr}. Moreover, the positive MF coupling  leads to the minimum chemical potential based on dispersion relation as: $\mu_{\min} = \mu_0= -(\frac{4m}{9\pi^2 \hbar^2})\, g^3/\delta g < 0,$  which establishes that the chemical potential is bounded below and is identical to the self trapped droplet condition. The soliton amplitude lies between zero and twice the constant background, taking the form,

\be
    \psi_{\pm} \left(x,t\right)=\frac{\sqrt{2m}}{3\pi \hbar}   \Big(\frac{g^{3/2}}{ \delta g}\Big)
    \left[1\pm\tanh\,( X/\xi)\right]\; exp\, [i\left(kx- \mu t/\hbar\right)], \label{eqn:8}
\ee

\begin{figure}[H]
\begin{center}
\includegraphics[width=6 cm,height=4.2 cm]{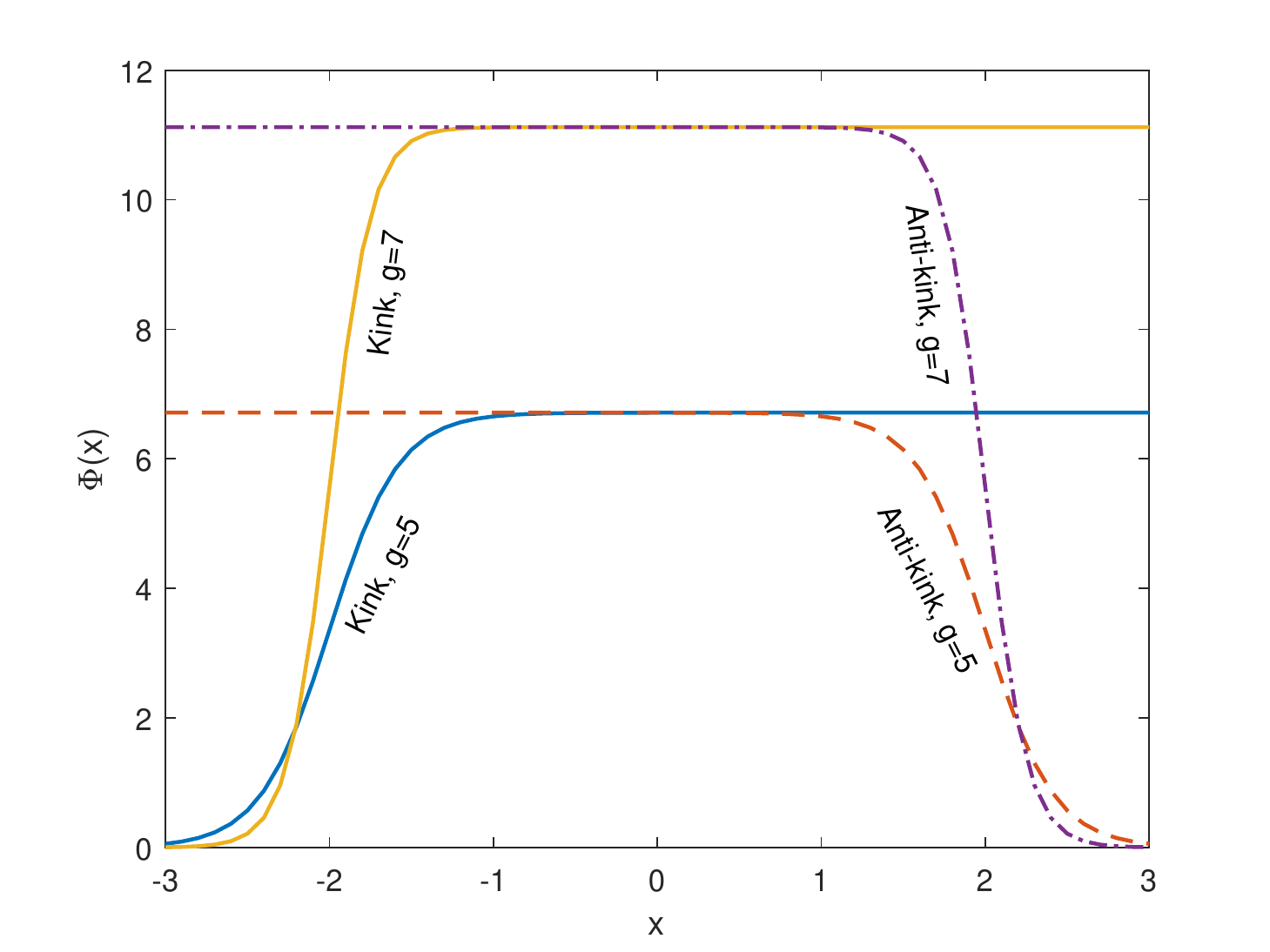}
\includegraphics[width=6 cm,height=4.0 cm]{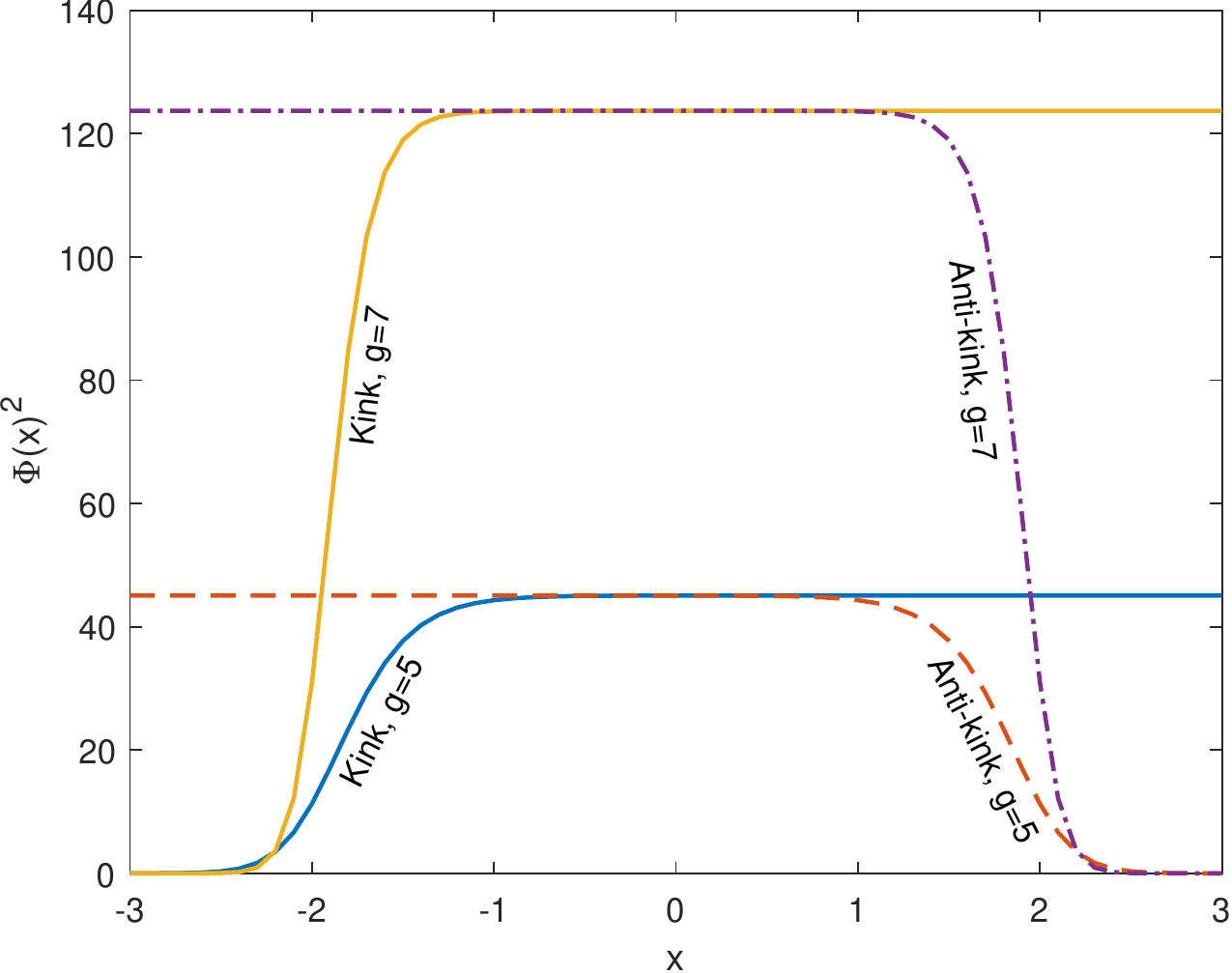}
\end{center}
\caption{(Color online) The variations of envelope profile and density are shown as function of position ($x$) for different values of $g$ (yellow: 5, blue: 7) with $\delta g =0.5$ at $t=1, v=2$ and $\hbar = m =1,$ interpolating smoothly the vacuum and droplet. The solid curve is for kink, while the dashed one is the anti-kink. }
\label{fig1}
\end{figure}
where $\xi = \frac{3\pi\hbar^2}{m} ( \delta g/ g^3)^{1/2}$. In Fig. \ref{fig1}, variations of the soliton profile and its density are depicted as a function of position for different values of BMF repulsion ($g$). The solid and dashed curves are for the two solitons travelling with equal velocity in opposite directions, ``$\psi_{s/d}=A\;\big[1\pm\text{tanh}\, \big(\frac{x\,\pm\, vt}{\xi}\big) \big]$". They can appear at the two boundaries of the droplet in a static configuration. The amplitude of the solitons increases by increase of the repulsive interspecies interaction. 
Interestingly,  $B = 0$ and  $B \rightarrow 0$ limits in the solution are not same. In case of $B=0,$ one obtains the constant background whereas $B \rightarrow 0$, profile has the constant background $\frac{\sqrt{2m}}{3\pi \hbar}\, (g^{3/2}/\delta g)$.

We now investigate the stability of these solitons through the Vakhitov-Kolokolov (VK) stability analysis \cite{vk}, which determines the `cost' of increasing the particle number density incrementally. For this purpose, we determine the number density,
\be
n=N/L=\frac{1}{L}\int_{-L/2}^{L/2} \left|\psi \right|^2~dx = \Big(\frac{4m}{9 \pi^2 \hbar^2}\Big)\, \frac{g^3}{\delta g},
\ee
the variation of number density with respect to the chemical potential determines the VK stability. For $L\rightarrow \infty$, VK stabilty criterion  results into,
\be
\frac{\partial n}{\partial \mu}=-\frac{1}{ \delta g}, \label{eqn:9}
\ee
 showing the stability of the kink-like soliton.

Before closing this section, it is worth pointing out that the kink-type solution with background is distinct from the grey soliton of NLSE with a complex profile, $\Psi_{GS} = \sqrt{\sigma}_0 \big(i sin \theta + cos \theta\, tanh\, (x\,cos \theta)\big)$.  In the present case, the analogous solution corresponding to grey soliton is: $\Phi (X) = A \,sin \theta + B\, cos\theta \,tanh \,\big( X cos \theta/\xi \big)$, with $\theta$ being the free parameter. Explicit calculation leads to, $A \,sin\theta = B\, cos\theta = \frac{\sqrt{2m}}{3\pi \hbar}\, (g^{3/2}/\delta g),$
   which yields the same kink-like profile given in Eq.(\ref{eqn:8}). In the following section, we discuss about the more general kink-like soliton and their properties.

\section{A General Kink-like Soliton} 

The solution of the amended GP equation in the form of NLSE with quadratic coupling can be obtained  through M\"obius or the so-called fractional transformation, connecting the solutions of solvable non-linear systems to the general ones under consideration \cite{p1}. The desired solution of ANLSE is taken as the general Pad\'e-type form \cite{pkp1, pkp2}, 
\be
\Phi(x)  = \frac{A + B \, f(x)}{1 + D \,f(x)}, \label{eqn:11}
\ee 
where $f(x)$ satisfy the elliptic function equation 
\be
 f'' (x) + a f^3 (x) + c f (x) = 0. \label{eqn:12}
\ee
In the present case, we consider the kink-type scenario with $f(x) = tanh\,(X/\xi)$. The emerging inter-connecting relationships, from the amended GP equation, between $A, B$ and $D$ are given as following:
\be
\begin{aligned}
&\frac{\hbar^2}{m\, \xi^2} (B - AD) - g_1 B^3 + g_2 B^2 D + \mu B D^2 = 0,\\
&\frac{\hbar^2}{m\, \xi^2} (B - AD) D - 3g_1 A B^2 + g_2 B^2 + 2g_2 ABD + \mu AD^2 + 2\mu BD = 0,\\
&g_1  B^3 + 3g_1 A^2 B - 2g_2 AB- g_2 A^2 D - g_2 B^2 D - 2\mu AD - \mu B - \mu B D^2= 0,\\
&g_1  A^3 - g_2 A^2 + 3g_1 A B^2 - 2g_2 ABD- g_2 B^2 - \mu A - \mu AD^2 -2 \mu B D = 0.
\label{eqn:13}
\end{aligned}
\ee
A straightforward but lengthy calculation leads to,
\be
(A - B)^2 = \frac{\mu}{g_1} (1-D)^2,\qquad
g_1 (A^2 + B^2) - g_2 (B + AD) - \mu(1+ D^2) = 0, \label{eqn:14}
\ee
explicitly, yielding the background $A$ and soliton amplitude $B$, in terms of $D$
\be
\begin{aligned}
 A &= B \pm (\mu/g_1)^{1/2} \;[1-D]\\
B &= \frac{1}{2g_2} \Bigl[ \Big(\pm 2 (\mu g_1)^{1/2} (D-1) + g_2 (D+1)\Big) \pm \Big( \big(\pm 2 (\mu g_1)^{1/2} (D-1) \\&+ g_2 (D+1)\big)^2 - 4 g_1 \big( \mp 2g_2 (\mu/g_1)^{1/2} D(1-D) - 4 \mu D \big)\Big)^{1/2} \Bigr],
\end{aligned} \label{eqn:15}
\ee
and the healing length  is given by 
 \be
\begin{aligned}
 \frac{1}{ \xi^2} = \Big(\frac{m}{\hbar^2}\Big)\, \frac{1}{(3-D^2)}\Big[3g_1 B^2 - 2g_2 BD - \mu D^2\Big].
 \end{aligned} \label{eqn:16}
\ee
The constant $D$ can be obtained by exploiting the relationships given in Eq.(\ref{eqn:15}) and Eq. (\ref{eqn:16}), together with the first two equations in Eq. (\ref{eqn:13}).

In our next section, we derive the energy and momentum for kink-like soliton. We also explain the BMF effect in energy density and show the procedure to counter the divergent background energy part. 

\section{ Energy and Momentum of the Kink-like Soliton}   
The ground state energy can be computed following the standard approach \cite{1r27}, with the Hamiltonian density,
\be
\mathcal{H}=\frac{\hbar^2}{2m}\left|\frac{\partial \Psi}{\partial x} \right|^2+\frac{1}{2}\, \delta g\left|\Psi\right|^4 - \frac{2}{3}\,\frac{\sqrt{2m}}{\pi \hbar}g^{3/2} \left|\Psi \right|^3.
    \label{eqn:17}
\ee
 We consider $\Psi(x, t) = \sqrt{\sigma}\, exp\,[i(kx - \omega t)]$, with  $\sqrt{\sigma} = A + B \, tanh\,(X/\xi)$ and use in the amended NLSE. As it will seen later, $\sigma$ is positive semi-definite for our ansatz solution. From the imaginary and real parts, one gets,
\be
\begin{aligned}
&\frac{\partial \sqrt{\sigma}}{\partial t} = - v\,\frac{\partial \sqrt{\sigma}}{\partial X} \qquad \quad\text{with}\quad v = \frac{\hbar k}{m},\\
&\frac{\hbar^2}{2m}\,\frac{\partial}{\partial X} \Bigl( \frac{\partial \sqrt{\sigma}}{\partial X} \Bigr)^2 = - \Big( \mu - \frac{\hbar^2 k^2}{m}\Big)\, \frac{\partial \sigma}{\partial X} + \frac{\delta g}{2} \;\frac{\partial \sigma^2}{\partial X} - \frac{2}{3}\; \frac{\sqrt{2m}}{\pi \hbar}g^{3/2}\, \frac{\partial\sigma^{3/2}}{\partial X}.
\end{aligned}
\label{eqn:18}
\ee
 As mentioned earlier, the density of grey soliton is constant  at asymptotic ends. In order to obtain the converging energy, constant background density has to be subtracted from the Hamiltonian (See, e.g., \cite{1r27}). However, for the kink-like case, inconsistent density distributions at the boundaries, lead to a non-trivial background subtraction. It is to be noted that the difference in densities, between two asymptotic ends, of the kink-like soliton is  $\frac{8m}{9\pi\hbar^2}( g^3/\delta g^2)$. We now consider the rest-frame with $k = 0$  and  integrate the above equation by taking the appropriate boundary densities to obtain
\be
\begin{aligned}
\frac{\hbar^2}{2m}\, \Bigl( \frac{\partial \sqrt{\sigma}}{\partial X} \Bigr)^2 = \frac{\delta g}{2} \sigma^2 - \frac{2}{3}\; \frac{\sqrt{2m}}{\pi \hbar}g^{3/2}\, \sigma^{3/2} - \mu\, \sigma +   \frac{32\,m^2}{81\pi^4\hbar^4}\Big(\frac{g^6}{\delta g^3}\Big) + \frac{8m}{9\pi^2\hbar^2} \,\mu\,\Big( \frac{g^3}{ {\delta g}^2}\Big),
\end{aligned} \label{eqn:19}
\ee
 the energy density in terms of $\sigma$, can be written in the following form 
\be
\mathcal{H}=\frac{\hbar^2}{2m}\Bigl(\frac{\partial \sqrt{\sigma}}{\partial X}\Bigr)^2+\frac{\delta g}{2}\,\sigma^2 - \frac{2}{3}\,\frac{\sqrt{2m}}{\pi \hbar}g^{3/2} \sigma^{3/2} - \mu \sigma.
    \label{eqn:20}
    \ee
Substituting Eq. (\ref{eqn:19}) in Eq. (\ref{eqn:20}),  energy for the kink-like soliton can be represented as:
\be
\begin{aligned}
\mathcal{E} =  \int_{- \infty}^{\infty}\, dX\; \mathcal{H} &= \int_{- \infty}^{\infty}\, dX\; \Bigl[g_1 \sigma^2 - \frac{4}{3} \;\frac{\sqrt{2m}}{\pi \hbar}g^{3/2}\, \sigma^{3/2} - 2 \mu\, \sigma \\
&+   \frac{32\,m^2}{81\pi^4\hbar^4}\Big(\frac{g^6}{\delta g^3}\Big) + \frac{8m}{9\pi^2\hbar^2} \,\mu\,\Big( \frac{g^3}{ {\delta g}^2}\Big)\Bigr].
\end{aligned} \label{eqn:21}
\ee
It is important to point out that the last two constant terms are the background energy which compensate the divergent terms in the energy density. One obtains the final expression of total energy, by exploiting the relationships achieved in Eq. (\ref{eqn:7}), as following
\be
\mathcal{E}= \frac{8\sqrt{2} }{27}\, \Big(\frac{m g^2}{\pi^3 \hbar^2}\Big)\, \Bigl(\frac{ g}{\delta g}\Bigr)^{5/2},  \label{eqn:22}
\ee
it is evident from above that the total energy is controlled by interaction parameters $\delta g$ and $g$.
The momentum  can be obtained by considering the  kink-like soliton in a  a box of finite length $L$ from ($-L/2$ to $L/2$). For solitonic profile $\Psi(x)$, one gets
\be
\begin{aligned}
\mathcal{P} = -\frac{i\hbar}{2}\,  \int_{-L/2}^{L/2} dx \Big[\Psi^{\star}\, \frac{\partial \Psi}{\partial x} -\Psi\, \frac{\partial \Psi^{\star}}{\partial x}\Big] = \hbar k\, A^2 \,\big[L - \xi\, tanh(L/2) \big].
\end{aligned}
\ee
As expected, the first term comes from the background $A\, exp\,[i(k x -\mu t/\hbar]$, whereas the second term arises from the soliton contribution. Interestingly, the two contributions have opposite sign which physically signifies that the background and solitonic profiles are moving in opposite directions. The soliton contribution in $L \rightarrow \infty$ limit yields
\be
\mathcal{P} = - \frac{2 \sqrt{2}}{3\pi}\,mv\, \Big(\frac {g}{\delta g}\Big)^{3/2} \equiv  M\,v,
\ee
where we used  $v = \frac{\hbar\,k}{m}$ and Eq. (\ref{eqn:7}) and $M = -\frac{2 \sqrt{2}}{3\pi}\,m\,(g/\delta g)^{3/2} $ is the effective mass akin to the earlier observed negative mass for the grey soliton in a trap \cite{malo4, kev3, pkp5}.
The energy of kink-like solion, in terms of effective mass, can be represented in the form
\be
\mathcal{E}= - \frac{4M}{9\pi^2 \hbar^2}\, \Bigl(\frac{ g^3}{\delta g}\Bigr).  \label{eqn:23}
\ee
Remarkably, on comparing the above equation with the lowest value of chemical potential for self-bound droplet $\mu_0 = -\frac{4m}{9\pi^2 \hbar^2}\,( g^3/\delta g)$, one obtains $\mathcal{E} = (M/m) \mu_0$. 

\section{Conclusion}
In conclusion, we have obtained the kink-like quantum soliton solutions in the quantum liquid, which smoothly interpolate between the normal phase and the flat-top droplet. They necessarily require a constant background of quantum nature. Remarkably, the background amplitude is exactly one-third of the uniform condensate amplitude. Evidently, these kink-like solitons are different from kink/anti-kink, dark, and grey solitons, as at one end, asymptotically they connect to the normal state. These dark and grey solitons  represent localized defects.  The chemical potential has a value $\mu_{0}$, which is identical to the condition for the self-trapped flat-top solution. The possibilty and nature of extended objects like, Ma and Akhmediev breathers \cite{brth1, brth2}, as well as rogue waves is worth studying for the droplets \cite{ brth3}. The collision of solitons and their role in the evaporation of the droplets is under investigation. 
\vskip 1cm

\ack {AS and PKP acknowledge the support from DST, India through Grant No.: DST/ICPS/QuST/Theme-1/2019/2020-21/01. Neeraj is thankful to DST, India for providing fellowship under the INSPIRE programme, the Grant No.: DST/INSPIRE Fellowship/2016/IF160177.}

\vskip 1.4cm

\end{document}